\documentclass[12pt]{article}
\begin{document}

\def\bd{\begin{displaymath}}\def\ed{\end{displaymath}}
\def\be{\begin{equation}}\def\ee{\end{equation}}
\def\bea{\begin{eqnarray}}\def\eea{\end{eqnarray}}
\def\nn{\nonumber}\def\lb{\label}

\def\a{\alpha}\def\b{\beta}\def\c{\chi}\def\d{\delta}\def\e{\epsilon}
\def\f{\phi}\def\g{\gamma}\def\h{\theta}\def\i{\iota}\def\j{\vartheta}
\def\k{\kappa}\def\l{\lambda}\def\m{\mu}\def\n{\nu}\def\o{\omega}\def
\p{\pi}\def\q{\psi}\def\r{\rho}\def\s{\sigma}\def\t{\tau}\def\u{\upsilon}
\def\vu{\varphi}\def\w{\varpi}\def\y{\eta}\def\x{\xi}\def\z{\zeta}

\def\D{\Delta}\def\F{\Phi}\def\G{\Gamma}\def\H{\Theta}\def\L{\Lambda}
\def\O{\Omega}\def\P{\Pi}\def\Q{\Psi}\def\S{\Sigma}\def\U{\Upsilon}\def\X{\Xi}

\def\lie{{\cal L}}\def\de{\partial}\def\na{\nabla}\def\per{\times}
\def\inf{\infty}\def\id{\equiv}\def\mo{{-1}}\def\ha{{1\over 2}}
\def\qu{{1\over 4}}\def\pro{\propto}\def\app{\approx}
\def\we{\wedge}\def\di{{\rm d}}\def\Di{{\rm D}}

\def\Ei{{\rm Ei}}\def\li{{\rm li}}\def\const{{\rm const}}\def\ex{{\rm e}}
\def\arcsh{{\rm arcsinh}}\def\arcch{{\rm arccosh}}
\def\arcth{{\rm arctanh}}\def\arccth{{\rm arccoth}}
\def\diag{{\rm diag}}

\def\gmn{g_{\m\n}}\def\ep{\e_{\m\n}}\def\ghmn{\hat g_{\m\n}}\def\mn{{\mu\nu}}
\def\dix{\int d^2x\ \sqrt{-g}\ }\def\ds{ds^2=}\def\sg{\sqrt{-g}}
\def\dhx{\int d^2x\ \sqrt{-\hat g}\ }\def\dex{\int d^2x\ e\ }

\def\tors#1#2#3{T_{#1#2#3}}\def\curv#1#2#3#4{R_{#1#2#3#4}}
\def\af{asymptotically flat }\def\hd{higher derivative }\def\st{spacetime }
\def\fe{field equations }\def\bh{black hole }\def\as{asymptotically }
\def\tran{transformations }\def\ther{thermodynamical }\def\coo{coordinates }
\def\bg{background }\def\gs{ground state }\def\bhs{black holes }
\def\sc{semiclassical }\def\hr{Hawking radiation }\def\sing{singularity }
\def\ct{conformal transformation }\def\cc{coupling constant }
\def\crel{commutation relations }
\def\ns{naked singularity }\def\gi{gravitational instanton }
\def\rep{representation }\def\gt{gauge transformation }
\def\cco{cosmological constant }\def\em{electromagnetic }
\def\ssy{spherically symmetric }\def\cf{conformally flat }
\def\cur{curvature }\def\tor{torsion }\def\ms{maximally symmetric }
\def\coot{coordinate transformation }\def\diff{diffeomorphisms }
\def\pb{Poisson brackets }\def\db{Dirac brackets }\def\ham{Hamiltonian }
\def\cd{covariant derivative }\def\dof{degrees of freedom }
\def\hdim{higher dimensional }\def\ldim{lower dimensional }
\def\dys{dynamical system }\def\cps{critical points }\def\dim{dimensional }
\def\sch{Schwarzschild }\def\min{Minkowski }\def\ads{anti-de Sitter }
\def\RN{Reissner-Nordstr\"om }\def\RC{Riemann-Cartan }\def\poi{Poincar\'e }
\def\KK{Kaluza-Klein }\def\pds{pro-de Sitter }\def\des{de Sitter }
\def\BR{Bertotti-Robinson }\def\MP{Majumdar-Papapetrou }
\def\GR{general relativity }\def\GB{Gauss-Bonnet }\def\CS{Chern-Simons }
\def\EH{Einstein-Hilbert }\def\EPG{extended \poi group }
\def\dpa{deformed \poi algebra }\def\psm{Poisson sigma model }

\def\PL#1{Phys.\ Lett.\ {\bf#1}}\def\CMP#1{Commun.\ Math.\ Phys.\ {\bf#1}}
\def\PRL#1{Phys.\ Rev.\ Lett.\ {\bf#1}}\def\AP#1#2{Ann.\ Phys.\ (#1) {\bf#2}}
\def\PR#1{Phys.\ Rev.\ {\bf#1}}\def\CQG#1{Class.\ Quantum Grav.\ {\bf#1}}
\def\NP#1{Nucl.\ Phys.\ {\bf#1}}\def\GRG#1{Gen.\ Relativ.\ Grav.\ {\bf#1}}
\def\JMP#1{J.\ Math.\ Phys.\ {\bf#1}}\def\PTP#1{Prog.\ Theor.\ Phys.\ {\bf#1}}
\def\PRS#1{Proc.\ R. Soc.\ Lond.\ {\bf#1}}\def\NC#1{Nuovo Cimento {\bf#1}}
\def\JP#1{J.\ Phys.\ {\bf#1}} \def\IJMP#1{Int.\ J. Mod.\ Phys.\ {\bf #1}}
\def\MPL#1{Mod.\ Phys.\ Lett.\ {\bf #1}} \def\EL#1{Europhys.\ Lett.\ {\bf #1}}
\def\AIHP#1{Ann.\ Inst.\ H. Poincar\'e {\bf#1}}\def\PRep#1{Phys.\ Rep.\ {\bf#1}}
\def\AoP#1{Ann.\ Phys.\ {\bf#1}}
\def\grq#1{{\tt gr-qc/\-#1}}\def\hep#1{{\tt hep-th/\-#1}}

\def\den{\left(1-{P_0\over\k}\right)}\def\dem{\left(1-{\y_0\over\k}\right)}
\def\by{\bar\y}\def\yq{{\y_1^2-\y_0^2}}\def\gk{\left(1+{\G\over\k}\right)}
\def\eps{\e_{ijk}}

\begin{titlepage}
%\begin{flushright}INFNCA-TH0202 \\ \end{flushright}
\vspace{.3cm}
\begin{center}
\renewcommand{\thefootnote}{\fnsymbol{footnote}}
{\Large \bf Two-dimensional gravity with an invariant energy scale
and arbitrary dilaton potential}
\vfill%\vskip 15mm%27.mm
{\large \bf {S.~Mignemi\footnote{email: smignemi@unica.it}}}\\
\renewcommand{\thefootnote}{\arabic{footnote}}
\setcounter{footnote}{0}
\vfill%\vskip 7mm%1cm
{\small
  Dipartimento di Matematica, Universit\`a di Cagliari,\\
Via Ospedale 72, 09124 Cagliari, Italy\\
\vspace*{0.4cm} INFN, Sezione di Cagliari\\
}
\end{center}
\vfill
\centerline{\bf Abstract}
\vfill

We investigate a model of two-dimensional gravity with arbitrary scalar
potential obtained by gauging a deformation of de Sitter or more general
algebras, which accounts for the existence of an invariant energy scale.
We obtain explicit solutions of the field equations and discuss their
properties.

\vfill
\end{titlepage}
Deformations of the \poi algebra allowing for the existence of a
fundamental length (energy) scale have recently been the subject
of several investigations \cite{dpa,MS}. Their interest is motivated
by the fact that theories of quantum gravity seem to imply the
existence of a minimal length scale of the order of the Planck
length \cite{Ga}.

One may wonder whether the framework of deformed \poi algebra
can be extended to include gravitational interactions, giving
rise in this way to an
effective modification of general relativity at small scales.
As is well known, a theory of gravity can be obtained in
general by gauging the spacetime symmetry group, and this approach
was adopted in \cite{Mi1} for constructing a two-dimensional
gravitational model invariant under the specific deformation of
\poi algebra introduced in ref.\ \cite{MS}.

In this letter we go a step further, extending the model of
\cite{Mi1} to the case where a dilaton potential is present \cite{Mi2}.
This class of two-dimensional theories includes between others
(anti-)\des \cite{IT}, or extended \poi models \cite{CJ}.
Finding suitable deformations of the algebras of \cite{Mi2}
compatible with the assumptions of \cite{MS} is not trivial,
because the Jacobi identities impose constraints on the form
of the deformed algebra.

\bigbreak
The \dpa of \cite{MS} is given in two dimensions by the \crel
\be\lb{algebra}
[P_a,P_b]=0,\quad [J,P_0]=P_1-{P_0P_1\over\k},\quad
[J,P_1]=P_0-{P_1^2\over\k},
\ee
where $P_a$ are the generators of translations and $J$ that of
boosts and $a=0,1$. Tangent space indices are lowered and raised
by the tensor $h_{ab}=\diag(-1,1)$. We also make use of the
antisymmetric tensor $\e_{ab}$, with $\e_{01}=1$.
The deformation parameter $\k$ has the dimension
of a mass and can be identified with the inverse of the Planck
length.
The algebra (\ref{algebra}) can be considered as an example of
nonlinear algebra \cite{Ike}.

A gravitational model which is invariant under the local symmetry
(\ref{algebra}) was studied in \cite {Mi1} and is described by a
lagrangian
\be\lb{lag'}
L=\y_2R+\y_aT^a+{\y_1\over\k}\y_a\,\o\we e^a,
\ee
where $R=\di\o$ is the curvature and $T^a=\di e^a+\e^a_{\ b}\,\o\we e^b$
the torsion and $\y_a$, $\y_2$ are a triplet of scalar fields in the
coadjoint \rep of the algebra. $\y_2$ is often called dilaton.

We want to generalize this two-dimensional model to include
dilaton-dependent potentials $\L(\y_2)$, analogous to those studied
in \cite{Mi2} in the Lorentz-invariant case. Simple
special cases were de Sitter gravity \cite {IT}, corresponding to
$\L=\l\y_2$,
or extended \poi gravity \cite{CJ}, corresponding to $\L=\l$.
In these models, the commutation relation between the generators
of translations $P_a$ is of the form \cite{Mi2}
\be\lb{pp}
[P_0,P_1]=\L(J),
\ee
where $\L$ is the same function as the potential, and $J$ is the
generator of boosts.

Unfortunately, it is not possible to simply change the momentum
commutation relation in (\ref{algebra}) to the form (\ref{pp})
because the set does not satisfy the generalized Jacobi identities
of nonlinear algebras \cite{Ike}.
The simplest generalization of the algebra (\ref{algebra})
is to admit a relation of the form $[P_0,P_1]=f(P_a)\L(J)$.
A simple calculation shows that the Jacobi identities are in fact
satisfied if $f(P_a)=\den^3$. Hence, we shall consider the
nonlinear algebra
\be\lb{alg}
[P_a,P_b]=\den^3\L(J),\quad [J,P_0]=P_1-{P_0P_1\over\k},\quad
[J,P_1]=P_0-{P_1^2\over\k}.
\ee
For $\k\to\inf$, the algebra reduces to that
obtained in \cite{Mi2} for the Lorentz-invariant limit.
The algebra (\ref{alg}) admits a Casimir invariant
\be\lb{Casimir}
C={P_1^2-P_0^2\over\den^2}+2\int\L(J)dJ.
\ee

In order to construct a gauge theory for this algebra, we
proceed as in \cite{Mi1} and adopt the formalism of Ikeda \cite{Ike}.
Given an algebra
with \crel $[T_A,T_B]=W_{AB}(T)$, one introduces gauge fields
$A^A$ and a coadjoint multiplet of scalar fields $\y_A$, which
under infinitesimal \tran of parameter $\x^A$ transform as
\bea\lb{trans}
&&\d A^A=\di\x^A+U^A_{BC}(\y)A^B\x^C,\cr
&&\d\y_A=-W_{AB}(\y)\x^B,
\eea
where $U^A_{BC}$ and $W_{AB}$ are functions of the fields $\y_A$
which satisfy
\bea
U^A_{BC}={\de W_{BC}\over\de\y_A}.
\eea

In our case $A=0,1,2$, and we identify $T^a$ with $P^a$ and $T^2$ with $J$.
In analogy with \cite{IT,CJ}, we can then also identify
$A^a$ with the zweibeins $e^a$ and $A^2$ with the spin
connection $\o$, and obtain a gauge theory for gravity in two
dimensions.
A gauge-invariant lagrangian can then be obtained with the methods
of \cite{Ike}, and reads
\be
L=\y_2R+\y_aT^a+\ha\dem^3\L(\y_2)\e_{ab}e^a\we e^b+
{\y_1\over\k}\y_a\,\o\we e^a,
\ee
with \fe
\bea\lb{feq1}
&&\di\y_0=\left(\y_1-{\y_1\y_0\over\k}\right)\o-\dem^3\L(\y_2)\,e^1,\cr
&&\di\y_1=\left(\y_0-{\y_1^2\over\k}\right)\o+\dem^3\L(\y_2)\,e^0,\cr
&&\di\y_2=-\left(\y_1-{\y_1\y_0\over\k}\right)e^0-\left(\y_0-
{\y_1^2\over\k}\right)e^1,\cr
&&T^0=\o\we{\y_1\over\k}\,e^0+{3\over\k}\dem^2\L(\y_2)\,e^0\we e^1,\cr
&&T^1=\o\we\left({\y_0\over\k}\,e^0+{2\y_1\over\k}\,e^1\right),\cr
&&R=-\dem^3\L'(\y_2)\,e^0\we e^1.
\eea

For solving the field equations it is useful to use the \psm
formalism.
We shall not give here the details of this formalism and refer the
reader to the vast literature on the subject \cite{stro}.
For our purposes it is sufficient to define a new set of variables
(target space coordinates),
\be\lb{tsc}
X_1=C(\y_A),\qquad X_2=\arcch{\y_0\over\sqrt\yq},\qquad X_3=-\y_2,
\ee
where $C(\y_A)$ is the Casimir invariant (\ref{Casimir}) written in terms
of the $\y_A$, $C(\y_A)=\dem^{-2}(\y_1^2-\y_0^2)+2\int\L(\y_2)\di\y_2$.
The transformation (\ref{tsc}) holds when the value of
the function $\g=C-2\int\L(\y_2)d\y_2$ is negative.
For positive values of $\g$, one must perform instead the transformation
\be
X_1=C(\y_A),\qquad X_2=\arcsh{\y_0\over\sqrt\yq},\qquad X_3=-\y_2.
\ee
For definiteness, in the following we shall consider the case $\g<0$.

The new variables satisfy the \crel
\be
[X_1,X_2]=[X_1,X_3]=0,\qquad [X_2,X_3]=1,
\ee
and in term of them, the $\y_a$ can be written as
\be
\y_0={\G\over\D}\cosh X_2,\qquad\y_1={\G\over\D}\sinh X_2,
\ee
where
\be
\G=\sqrt{\left|X_1+2\int\L(-X_3)\di X_3\right|},\qquad \D=1+{\G\over\k}\cosh X_2.
\ee
In these variables, the field equations take a very simple form
\cite{stro}:
\be
\di X_1=0,\quad\di X_2=-B_3,\quad\di X_3=B_2,\quad\di B_i=0,
\ee
whose solutions are given by
\be
X_1=\const,\quad B_1=2\di\f,
\ee
$\f$, $X_2$ and $X_3$ being arbitrary functions.
One can then go back to the original variables. Defining
$X_1=c$, where $c$ is the constant value of $C$, $X_2=\h$,
$X_3=-\y$, one obtains
\bea
&&e^0=-\D\left[{1\over\G}\sinh\h\di\y+\G\left(\cosh\h-{\G\over\k}
\right)\di\f)\right],\cr
&&e^1=\D\left[{1\over\G}\cosh\h\di\y+\G\sinh\h\di\f)\right],\cr
&&\o=\di\h+\L\di\f
\eea
where, in terms of the new variables, the functions $\G$ and $\D$ read
$\G=\sqrt{|c-2\int\L\di\y|}$ and $\D=1+\G\cosh\h/\k$.

One is still free to choose a gauge. The most interesting
choices are $\h=0$ or $\h=\f$.
The first choice may be considered as the ground state of
the model and yields
\be
e^0=-\G\left(1-{\G^2\over\k^2}\right)\di\f,\qquad e^1={1\over\G}\gk\di\y,
\qquad\o=\L\di\f,
\ee
with
\be T^0={3\L\over\k}\gk\di\f\we\di\y,\qquad T^1=0,\qquad
R=-\L'\di\f\we\di\y.
\ee
In particular, for $\L=\l\y_2$, one obtains a deformation of
(anti-)de Sitter space with constant curvature, but nonvanishing
torsion.

In the second case,
\bea
&&e^0=-\D\left[{1\over\G}\sinh\h\di\y+\G\left(\cosh\h-{\G\over\k}
\right)\di\h\right],\cr
&&e^1=\D\left[{1\over\G}\cosh\h\di\y+\G\sinh\h\,\di\h\right],\cr
&&\o=(1+\L)\,\di\h,
\eea
and the components of the torsion and the curvature are
\bea
&&T^0=\left(-{1+\L\over\k}\sinh^2\h+{3\over\k}\L\D\right)\di\h\we\di\y,\cr
&&T^1={1+\L\over\k}\sinh\h\cosh\h\ \di\h\we\di\y,\cr
&&R=-\L'\,\di\h\we\di\y.
\eea

The solutions corresponding to $\g>0$ can be obtained from
the previous ones by simply
interchanging $\sinh\h$ with $\cosh\h$ everywhere.

For a discussion of the properties of the solutions, it may be
useful to define a line element $ds^2=h_{ab}e^ae^b$, although
this is not gauge invariant. The possibility of defining a
gauge invariant metric will be discussed elsewhere \cite{GKSV}.
Alternatively, one may interpret the lack of invariance
assuming that Planck-energy particles experience different
metrics depending on their state of motion.
One has
\be
ds^2=\D^2\left[{\di\y^2\over\G^2}-\G^2\left(1-2{\G\over\k}\cosh\h
+{\G^2\over\k^2}\right)\di\h^2+2{\G\over\k}\sinh\h\,\di\y\,\di\h\right]
\ee
for $\g<0$, or
\be
ds^2=\U^2\left[{-\di\y^2\over\G^2}+\G^2\left(1+2{\G\over\k}\sinh\h
-{\G^2\over\k^2}\right)\di\h^2+2{\G\over\k}\cosh\h\,\di\y\,\di\h\right]
\ee
for $\g>0$, with $\U=1+\G\sinh\h/\k$.

The discussion of the properties of the solutions is in general
quite complicated. First of all, we notice that the volume element
is $e^0\we e^1=\D^3\di\y\we\di\h$, which is positive definite when $\g<0$.
In this case, the only singularities (at least for well-beaheved $\L$)
are horizons occurring at the zeroes of $\G$.
When $\g>0$, instead, torsion singularities can occur at
$\sinh\h=-\k/\G$, where $\U$ vanishes, and the discussion of the solutions
becomes more intricated.

It can be interesting to consider as an example the case $\L=\l\y_2$,
which corresponds to de Sitter ($\l<0$) or anti-de Sitter ($\l>0$) in the
undeformed limit. Let us start from the case $\g<0$. In this case
the coordinate $\y$ is spacelike, and the solutions are
time-dependent deformations of the static \des or \ads solutions,
depending on the sign of $\l$.
When $\l$ and $c$ have the same sign, a horizon is present
at $\y=\sqrt{c/\l}$, but curvature and torsion are regular everywhere.

For $\g>0$, the coordinate $\y$ is timelike when $\U>0$, and the
solutions can
be interpreted as space-dependent deformations of the two-dimensional
cosmological \des or \ads solutions (see for example \cite{CM}).
Again a horizon is present when $\l$ and $c$ have the same
sign, but now also a singularity of the torsion occurs where $\U=0$.

A more accurate discussion of the properties of the solutions
would require the definition of a gauge-independent metric or
an interpretation of its gauge dependence, which we leave for
further investigations.
\bigbreak

We have investigated a model of gravity in two dimensions,
invariant under a deformation of the (anti)-de Sitter algebra
or of more general algebras corresponding to dilaton-dependent
potentials. We have defined the action and obtained the general
solutions of the field equations. All solutions imply the
presence of nontrivial torsion.

To conclude, we notice that a deformation of the (anti-)de Sitter
algebra compatible with the algebra of ref.\ \cite{MS}
in the $\l\to0$ limit
can be obtained also in four dimensions, using
a procedure analogous to the one adopted above.
The deformed algebra reads
\bea
&[P_i,P_j]=\l\den^2\left(\eps M_k+{1\over\k}(N_iP_j-N_jP_i)
\right),&\nn\\
&[P_0,P_j]=\l\den^3 N_j,\quad[M_i,P_j]=\eps
P_k,\quad[M_i,P_0]=0,&\nn\\
&[N_i,P_j]=\d_{ij}P_0-{P_iP_j\over\k},\quad[N_i,P_0]=
P_i-{P_0P_i\over\k},&\nn\\
&[N_i,N_j]=\eps M_k,\quad[M_i,N_j]=\eps N_k,\quad[M_i,M_j]=
\eps M_k,&\nn
\eea
where $i,j,...=1,2,3$, $N_i=M_{0i}$, $M_k=\ha\eps M_{ij}$.

Unfortunately, however, an action principle for nonlinear gauge
theories is not known in four dimensions.
Nevertheless, we believe that our low-dimensional model
can give some hints for addressing this problem, and also other
open questions, as the definition of a gauge-invariant metric
and/or a suitable coupling for point particles.

\end{document}